# Antiferromagnetic criticality at a heavy-fermion quantum phase transition

W. Knafo[1,2]*, S. Raymond[1], P. Lejay[3] and J. Flouquet[1]

1 Service de Physique Statistique, Magnétisme et Supraconductivité, Institut Nanosciences et Cryogénie, Commissariat à l'Energie Atomique, 17 rue des Martyrs, 38054 Grenoble cedex 9, France,

2 Laboratoire National des Champs Magnétiques Intenses, UPR 3228, CNRS-UPS-INSA-UJF, 143 Avenue de Rangueil, 31400 Toulouse, cedex 4, France,

3 Institut Néel, CNRS/UJF, 25 avenue des Martyrs, BP 166, 38042 Grenoble cedex 9, France.

*e-mail: knafo@lncmp.org.

**The interpretation of the magnetic phase diagrams of strongly correlated electron systems remains controversial. In particular, the physics of quantum phase transitions, which occur at zero temperature, is still enigmatic. Heavy-fermion compounds aretextbook examples of quantum criticality, as doping, or the application of pressure or a magnetic field can lead to a quantum phase transition between a magnetically ordered state and a paramagnetic regime. A central question concerns the microscopic nature of the critical quantum fluctuations. Are they antiferromagnetic or of local origin? Here we demonstrate, using inelastic neutron scattering experiments, that the quantum phase transition in the heavy-fermion system $Ce_{1-x}La_xRu_2Si_2$ is controlled by fluctuations of the antiferromagnetic order parameter. At least for this heavy-fermion family, the Hertz–Millis-Moriya spin fluctuation approach seems to be a sound basis for describing the quantum antiferromagnetic–paramagnetic instability.**

Inelastic neutron scattering experiments were decisive for determining the role of spin fluctuations in itinerant 3d-electron magnets[1], confirming the pertinence of the spin-fluctuation theories developed for these systems [2]. These theories are based on the assumption that the Fermi energy scale is much bigger than the magnetic energy scales, which is indeed the case for 3d-electron systems. The case of heavy-fermion itinerant magnets is more complex: in these systems, the f-electrons have an itinerant character and the Fermi energy scale is strongly renormalized by single-site f-c hybridization. The description of heavy-fermion materials is made difficult because the Fermi temperature, related to the high effective mass of the so-called heavy electrons, is reduced to the order of the magnetic energy scales.

Quantum criticality, that is, the critical properties at a quantum instability, is at the heart of the heavy-fermion problem, and more generally of the physics of a large class of strongly correlated electronic systems [3-6]. The generic phase diagram of heavy fermions, shown schematically in Fig. 1, depends on an external parameter _, such as pressure, doping or a magnetic field. At $\delta = \delta_c$ and $T = 0$, a second-order quantum phase transition, also called a quantum critical point, separates a paramagnetic Fermi-liquid regime from the magnetically ordered state, which is often antiferromagnetic. The Fermi liquid and the antiferromagnetic regimes occur below the Fermi temperature $T^*$ and the Néel temperature $T_N$, respectively, which are both expected to vanish at the quantum critical point. In the Hertz-Millis-Moriya (HMM) spin-fluctuation theory of quantum phase transitions [7-9], fluctuations of the antiferromagnetic moment, that is of the order parameter, control the phase transition. In this theory, critical antiferromagnetic fluctuations govern the Fermi-liquid regime and their intensity diverges at the quantum critical point. Experimentally, a `non-Fermi liquid' is observed in the vicinity of most heavy-fermion quantum critical points [10] and is believed to result from the effects of temperature on the critical fluctuations. Despite encouraging attempts [11,12], standard HMM theories fail to describe quantitatively the anomalous properties of the non-Fermiliquid regime [10]. Recently, a new theoretical approach, based on the concept of local critical magnetic fluctuations, has been proposed by Coleman and Si and co-workers [13-16] (CS). This `local' scenario was originally proposed to explain inelastic neutron scattering data obtained on the critical heavy-fermion compound $CeCu_{5.9}Au_{0.1}$ (refs 17, 18). `Conventional' spin fluctuation models were also used 20 years ago to describe inelastic neutron scattering experiments on the parent compound CeCu$_6$ (refs 19, 20), as well as macroscopic measurements on other heavy-fermion and intermediate valence systems [21]. However, a complete microscopic study and comparison of the spin fluctuations in all regions of a heavy-fermion phase diagram, including the magnetically ordered phase, has never been carried out.

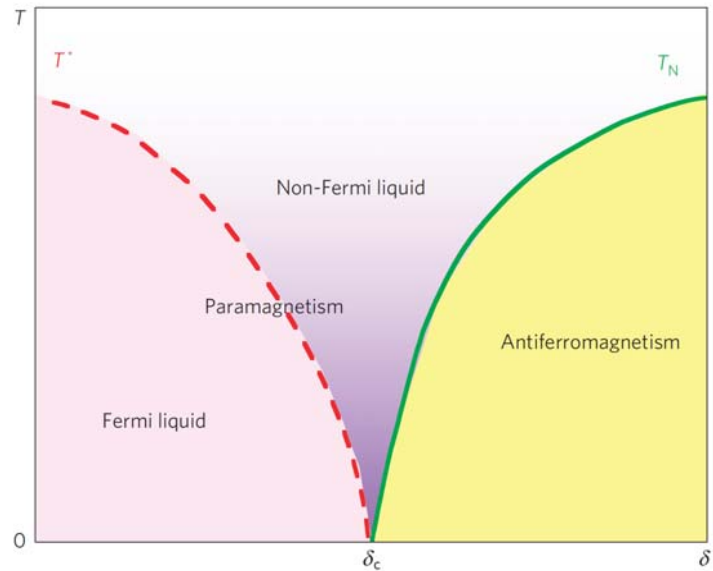

**Figure 1** : **Schematic phase diagram of heavy-fermion systems.** $T^*$ is the energy scale of the Fermi-liquid regime and $T_N$ is the Néel temperature, characteristic of the antiferromagnetic phase. $\delta$ corresponds to an adjustable parameter, such as chemical doping, pressure or a magnetic field.

Here, we present a systematic experimental investigation of the magnetic fluctuations in the heavy-fermion system Ce$_{1-x}$La$_x$Ru$_2$Si$_2$. This is the first direct comparison between the magnetic fluctuations from all parts of a heavy-fermion phase diagram, using a microscopic probe. Inelastic neutron scattering experiments were carried out for a wide range of temperatures and La concentrations x, in both the antiferromagnetic and the paramagnetic sides of the quantum phase transition and at two different momentum transfers Q. From our data, we demonstrate that, for the system Ce$_{1-x}$La$_x$Ru$_2$Si$_2$, quantum criticality is of an antiferromagnetic nature and can be described using a conventional HMM-like scenario [7-9], but not within the CS `local' scenario [13-16]. However, no collapse of the characteristic lowest temperature is observed, suggesting that the quantum phase transition may be weakly first order.

# Magnetic fluctuations in the heavy-fermion Ce$_{1-x}$La$_x$Ru$_2$Si$_2$

The archetypal heavy-fermion material CeRu$_2$Si$_2$ is a paramagnet characterized by strong Ruderman-Kittel-Kasuya-Yosida (RKKY) antiferromagnetic fluctuations, which lead to maxima of the dynamical magnetic susceptibility at the incommensurate wave vectors **k**$_1$ = (0.31,0,0), **k**$_2$ = (0.31,0.31,0) and **k**$_3$ = (0,0,0.35) in the low-energy spectra obtained by inelastic neutron scattering [12]. At wave vectors sufficiently far from k$_1$, k$_2$ and k$_3$, magnetic fluctuations persist and can be considered as the signature of a local (or single-site) Kondo effect. Its peaked spectrum in momentum transfer Q space observed by inelastic neutron scattering experiments indicates that CeRu$_2$Si$_2$ is close to the onset of long-range magnetic ordering. Indeed, appropriate chemical doping (with La, Ge or Rh) and/or magnetic field tunings can favour one particular interaction and establish antiferromagnetic long-range ordering with either k$_1$, k$_2$ or k$_3$ wave vectors. Elastic Bragg peaks have been detected in these antiferromagnetic states [22-25]. The proximity of CeRu$_2$Si$_2$ to paramagnetic_antiferromagnetic quantum phase transitions is demonstrated by the increase on cooling of the specific heat divided by temperature C$_P$=T (ref. 26) and of the electronic Grüneisen parameter [27]. A strong enhancement of the Sommerfeld coefficient $\gamma = (C_p/T)_{T \to 0}$ is also observed at the critical compound Ce$_{0.925}$La$_{0.075}$Ru$_2$Si$_2$ (ref. 11), indicating the presence of a quantum critical regime. The x-T phase diagram of Ce$_{1-x}$La$_x$Ru$_2$Si$_2$ is composed of a paramagnetic regime, for $x < x_c$ = 7.5 %, where a strongly renormalized Fermi liquid is reached at low temperatures, and of an antiferromagnetic phase, for $x > x_c$, where long-range magnetic order occurs with the incommensurate wave vector k$_1$ (refs 22) (La doping can be considered as equivalent to a `negative pressure' [28]). Ce$_{1-x}$La$_x$Ru$_2$Si$_2$ is an ideal system for studying quantum criticality because of its relatively simple structure (tetragonal symmetry) and its accessible magnetic energy scales, typically of the order of a few kelvin. Owing to the crystal-field splitting of the $J = 5/2$ sextuplet [29], magnetic anisotropy is of the Ising kind and the static antiferromagnetic moments, as well as the low-energy magnetic fluctuations, are aligned along the easy direction c. High-quality crystals and `relatively moderated' effective masses permit a complete determination of the Fermi surface and confirmed the picture of itinerant magnetism for CeRu$_2$Si$_2$. Its Fermi surface has been well reproduced by band-structure calculations with itinerant 4f electrons [30,31]. CeRu$_2$Si$_2$ is one of the rare heavy-fermion systems for which the Fermi surface has been almost fully determined by de Haas_van Alphen measurements [30,31], and quantum oscillations have also been recently reported for La-doped CeRu2Si2 compounds [32].

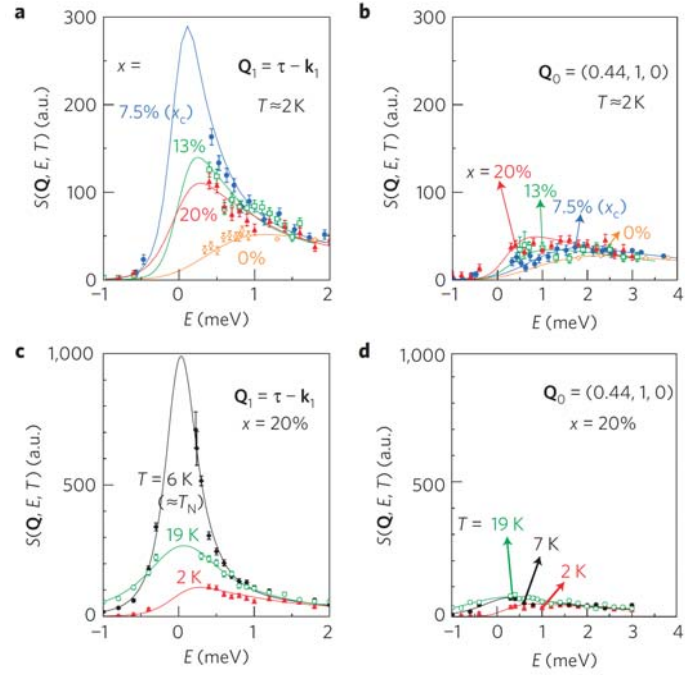

**Figure 2 : Inelastic neutron scattering spectra of Ce1-$x$La$x$Ru2Si2. a–d**, Neutron scattering intensity $S(\mathbf{Q},E,T)$ at the antiferromagnetic momentum transfer $\mathbf{Q}_1$ (**a,c**) and at the local momentum transfer $\mathbf{Q}_0$ (**b,d**), for various temperatures $T$ (**c,d**) and lanthanum concentrations $x$ (**a,b**). The lines are fits to the data using equations (1) and (2). The errors bars correspond to the square root of the number of neutron counts.

Figure 2 shows the spectra obtained by inelastic neutron scattering for Ce1-xLaxRu2Si2 for various dopings, temperatures and wave vectors. Systematic measurements were made on the compounds of concentrations $x$ = 0, 7.5, 13, and 20 % at two momentum transfers Q1 and Q0 (earlier studies of the magnetic fluctuations were limited to the paramagnetic phase [33-35]). $\mathbf{Q}_1 = (0.69,1,0)$ corresponds to the wave vector $\mathbf{k}_1 = \boldsymbol{\tau} - \mathbf{Q}_1$, $\boldsymbol{\tau} = (110)$ being the position of a structural Bragg peak. It is characteristic of the antiferromagnetic fluctuations with the wave vector $\mathbf{k}_1$. $\mathbf{Q}_0 = (0.44,1,0)$ is characteristic of uncorrelated magnetic fluctuations, being far from the different antiferromagnetic vectors k1, k2 and k3. Magnetic spectra at Q0 can be considered as the signature of local magnetic fluctuations, which are controlled by the single-site Kondo effect. We note that `local' is used to qualify a `single-site' wave vector q-independent phenomenon, as opposed to an `inter-site' q-dependent one, whereas `localized' is used to characterize the nature of the f-electrons, in opposition with `itinerant'. For the two momentum transfers $Q_1$ and $Q_0$, the x-dependence of the magnetic fluctuations in the limit $T \to 0$, that is, when the quantum phase transition at $x_C$ is crossed (Fig. 2a,b), looks very similar to the T-dependence of the magnetic fluctuations for $x > x_C$, that is, when the classical phase transition at $T_N$ is crossed (Fig. 2c,d). Below, we describe with more details the data plotted in these graphs. In Fig. 2a,b, the low-temperature ($T \approx 2$ K) spectra measured at the `antiferromagnetic' momentum transfer $Q_1$ and at the `local' momentum transfer $Q_0$, respectively, are shown for $x = 0, 7.5$ ($x_c$), 13, and 20 %. Figure 2a shows a strong enhancement of the low-energy antiferromagnetic fluctuations at $x_C$ and $T \to 0$, whereas Fig. 2b indicates that the intensity of the local fluctuations (at $Q_0$) increases monotonously with x, with no peculiar feature at $x_C$. In Fig. 2c, spectra at the antiferromagnetic momentum transfer $Q_1$ are given for the antiferromagnetic compound of concentration $x = 20$ %, at temperatures above and below $T_N$. Spectra at the local momentum transfer $Q_0$ are given in Fig. 2d for the same compound. These plots indicate that the low-energy antiferromagnetic fluctuations at $Q_1$ have a maximal intensity at the Néel temperature $T_N \approx 6$, but also that the intensity of local magnetic fluctuations at $Q_0$ varies monotonously with T. Consequently, we can infer that both the quantum phase transition, at $x_C$ and $T \to 0$ and the classical (or thermal) phase transition, at $T_N$ and $x = 20$ %, are characterized by an enhancement of the antiferromagnetic fluctuations at $k_1$ and by no noticeable change in the local magnetic fluctuations. In the following, we present a quantitative analysis of the data presented in Fig. 2 and discuss the microscopic nature of quantum criticality in $Ce_{1-x}La_xRu_2Si_2$.

## Static susceptibility and relaxation rate

Knowing that the neutron scattering intensity $S(\mathbf{Q},E,T)$ is related to the imaginary part of the dynamical susceptibility $\chi''(\mathbf{Q},E,T)$ by:

$$S(\mathbf{Q},E,T) = \frac{1}{\pi} \frac{1}{1-e^{-E/k_BT}} \chi''(\mathbf{Q},E,T) \, , (1)$$

the spectra presented above were fitted using a quasi-elastic Lorentzian line shape of the form:

$$\chi(\mathbf{Q},E,T) = \chi'(\mathbf{Q},E,T) + i\chi''(\mathbf{Q},E,T) = \frac{\chi'(\mathbf{Q},T) \times \Gamma(\mathbf{Q},T)}{\Gamma(\mathbf{Q},T) - iE}, \quad (2)$$

where $\chi'(\mathbf{Q},T)$ is the real part of the static susceptibility and $\Gamma(\mathbf{Q},T)$ is the relaxation rate.

In Fig. 3a,b, the parameters $\chi'(\mathbf{Q},T)$ and $\Gamma(\mathbf{Q},T)$ extracted at the antiferromagnetic momentum transfer $Q_1$ are plotted as a function of T for $x = 0$, 7.5 ($x_c$), 13, and 20 %. Maxima of $\chi'(\mathbf{Q}_1,T)$ and minima of $\Gamma(\mathbf{Q}_1,T)$ are observed at $x_c$ and $T \to 0$, and also at $T_N$ for $x = 13$ and 20 %. This confirms that low-energy antiferromagnetic fluctuations at the wave vector $k_1$ are maximal at both the quantum ($x_c, T \to 0$) and classical ($x > x_c, T_N$) phase transitions of the phase diagram of $Ce_{1-x}La_xRu_2Si_2$. The fact that similar saturated values, a minimum of $\Gamma(\mathbf{Q}_1, T) \approx 2$ K and a maximum of $\chi'(\mathbf{Q}_1, T) \approx 1500$ arb. units, are reached when $T_N(x)$ is crossed (at constant T or x) indicates that the saturation may have an intrinsic origin.

In Fig. 3d,e, the parameters $\chi'(\mathbf{Q},T)$ and $\Gamma(\mathbf{Q},T)$ extracted at the local momentum transfer $Q_0$ are plotted as a function of T for $x = 0$, 7.5 ($x_c$), 13, and 20 %. These plots show no singularity in the local magnetic fluctuations at the transition to the antiferromagnetic phase. Instead, a continuous increase of $\chi'(\mathbf{Q}_0,T)$ with x and a continuous decrease of $\Gamma(\mathbf{Q}_0,T)$ with x are obtained across the quantum phase transition ($x_c, T \to 0$). A continuous decrease of $\chi'(\mathbf{Q}_0,T)$ with T and a continuous increase of $\Gamma(\mathbf{Q}_0,T)$ with T are obtained across the classical phase transition ($T_N$, $x > x_c$). These data indicate that local Kondo-type fluctuations persist in the antiferromagnetic phase, which means the f –electrons still have a strong itinerant character in this regime.

As shown in Fig. 3c,f, the product $\chi'(\mathbf{Q},T) \times \Gamma(\mathbf{Q},T)$ is, within error, independent of x and T in the window [$x < 20$ %, $T < 20$ K], having similar average values for $Q_1$ and $Q_0$ (= 2950 and 3250 arbitrary units, respectively). This is in agreement with Kuramoto's description of paramagnetic heavy-fermion systems, where $\chi'(\mathbf{Q}) \times \Gamma(\mathbf{Q})$ is Q-independent [20,33,35]. Furthermore, our results suggest that such a model could be extended to the whole phase diagram of heavy-fermion systems, $\chi'(\mathbf{Q},T) \times \Gamma(\mathbf{Q},T)$ having a unique value for both the paramagnetic regime and the antiferromagnetically ordered phase of $Ce_{1-x}La_xRu_2Si_2$.

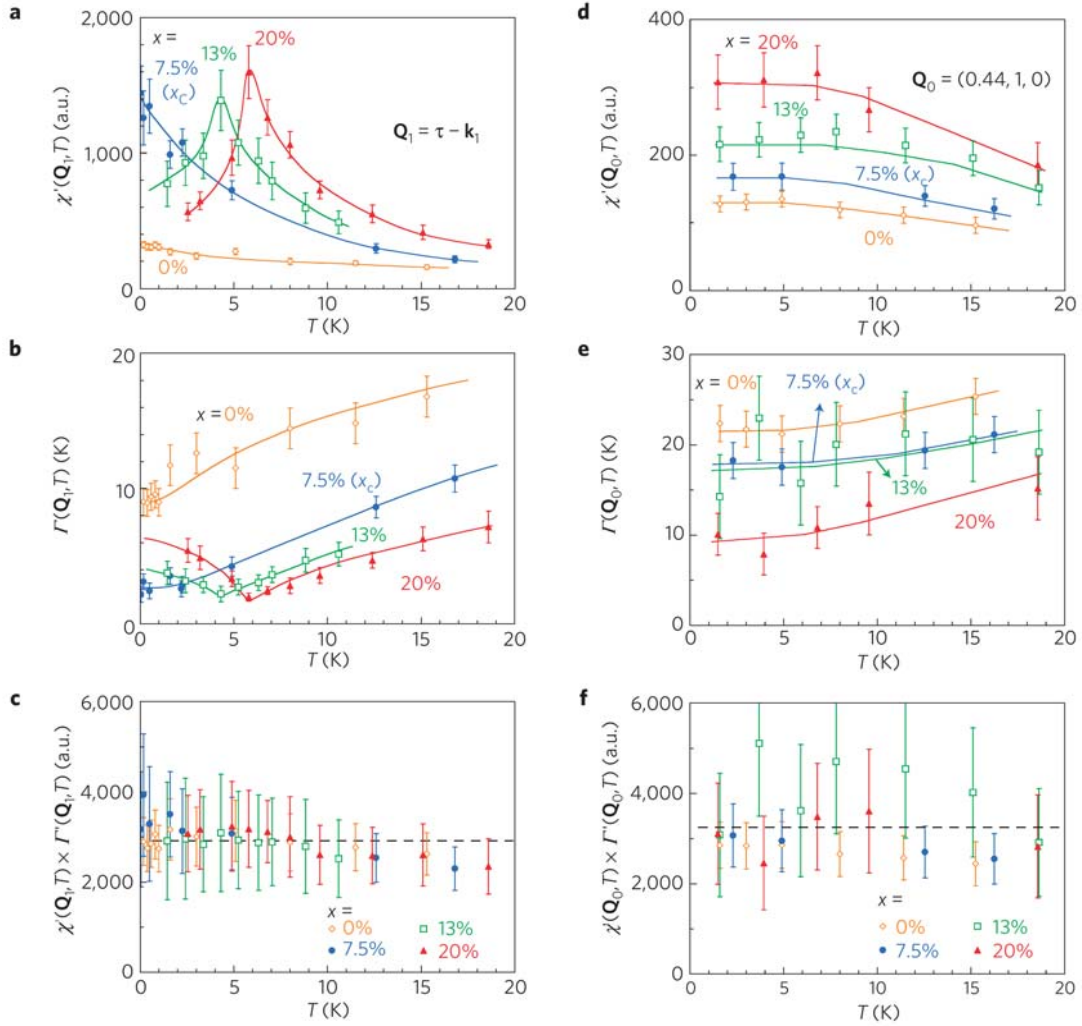

**Figure 3 : Static susceptibility and relaxation rate at antiferromagnetic and local wave vectors of $Ce_{1-x}La_xRu_2Si_2$. a–c**, Temperature variations of the static susceptibility $\chi'(\mathbf{Q}_1,T)$ (**a**), of the relaxation rate $\Gamma(\mathbf{Q}_1,T)$ (**b**) and of their product $\chi'(\mathbf{Q}_1,T) \times \Gamma(\mathbf{Q}_1,T)$ (**c**) at the antiferromagnetic momentum transfer $\mathbf{Q}_1$, for the lanthanum contents $x = 0$, 7.5 ($x_c$), 13 and 20 %. **d–f**, Temperature variations of the static susceptibility $\chi'(\mathbf{Q}_0,T)$ (**d**), of the relaxation rate $\Gamma(\mathbf{Q}_0,T)$ (**e**) and of their product $\chi'(\mathbf{Q}_0,T) \times \Gamma(\mathbf{Q}_0,T)$ (**f**) at the local momentum transfer $\mathbf{Q}_0$, for $x = 0$, 7.5 ($x_c$), 13 and 20 %. The solid lines are guides to the eyes and, the dotted lines show the best fits to $\chi'(\mathbf{Q}_1,T) \times \Gamma(\mathbf{Q}_1,T)$ and $\chi'(\mathbf{Q}_0,T) \times \Gamma(\mathbf{Q}_0,T)$ by a constant. The errors bars come from the numerical fits to the peak line shape described by equations (1) and (2).

The x-T phase diagram in Fig. 4 summarizes the magnetic energy scales and the magnitude of the real part of the static susceptibility $\chi'(\mathbf{Q}_1,T)$ determined by neutron scattering for $Ce_{1-x}La_xRu_2Si_2$. The variation with La doping x of the magnetic order parameter $M_0(\mathbf{k}_1)$ (extrapolated to $T = 0$ K) of the antiferromagnetic phase (from ref. 22) is also shown on top of the phase diagram. For $x > x_c$, the Néel temperature $T_N(x)$ delimits the antiferromagnetic phase, reaching 6 K at $x = 20$ % (ref. 22). The temperatures $T_1 \underset{T \to 0}{=} \Gamma(\mathbf{Q}_1,T)/k_B$ and $T_0 \underset{T \to 0}{=} \Gamma(\mathbf{Q}_0,T)/k_B$ are the characteristic energy scales of the antiferromagnetic fluctuations with the wave vector $k_1$ and of the local magnetic fluctuations, respectively. The minimum of $T_1$ indicates that the low-energy antiferromagnetic fluctuations are maximum at the quantum phase transition at $x_c$. As can be seen in the phase diagram, for $x = 13$ and 20 %, $T_1$ is roughly equal to the Néel temperature $T_N$. The absence of a minimum in the x-variation of $T_0$ signifies that the local fluctuations are not enhanced at the quantum phase transition. Knowing that $T_0$ can be interpreted as the Kondo temperature $T_K$ of the system, its x-dependence indicates that La doping reduces the strength of the Kondo effect and leads to an increase of the localized character of the f -electrons. These results are compatible with the Doniach picture [36], where a progressive localization of the f-electrons, induced by increasing x in $Ce_{1-x}La_xRu_2Si_2$, is accompanied by the development of RKKY interactions, which ultimately leads to long-range magnetic ordering.

## HMM-like order-parameter-fluctuations scenario

Our data support the scenario where antiferromagnetic fluctuations with wave vector $k_1$ are responsible for the quantum phase transition of $Ce_{1-x}La_xRu_2Si_2$, because their intensity is maximal at $(x_c, T \to 0)$. On the contrary, local magnetic fluctuations are not the driving force, because their intensity varies monotonically across the quantum phase transition. Theoretically, at a second-order quantum phase transition a divergence of the critical fluctuations is expected, accompanied by a vanishing of the associated characteristic temperature. However, instead of vanishing at $x_c$, the antiferromagnetic energy scale $T_1$ reaches a minimum and, instead of the divergence expected from the theory, there is a saturation of the susceptibility $\chi'(\mathbf{Q}_1, T \to 0)$ at the quantum phase transition (see also refs 33 and 37). Sample imperfections could be the reason for this saturation. Alternatively, the fact that similar minima of $\Gamma(\mathbf{Q}_1,T) \approx 2$ K and maxima of $\chi'(\mathbf{Q}_1,T) \approx 1500$ arb. units are obtained in crossing both $x_c$ and $T_N$ favours an intrinsic low-energy cutoff. This saturation may suggest that the quantum phase transition between the paramagnetic and antiferromagnetic phases at $x_c$ is first order (similar effects were reported for the itinerant magnets MnSi and $Sr_{1-x}Ca_xRuO_3$; ref. 38). This conclusion may be supported by the observation of tiny ordered moments in pure $CeRu_2Si_2$, which is assumed to be in a paramagnetic ground state39.

Validation of a conventional HMM-like scenario, in which order-parameter fluctuations are critical, requires a maximum of the antiferromagnetic order-parameter fluctuations at the quantum phase transition, together with a monotonous evolution of the magnetic fluctuations at the local momentum transfers. This is clearly the case for $Ce_{1-x}La_xRu_2Si_2$, for which the antiferromagnetic fluctuations with wave vector $k_1$ and not the local magnetic fluctuations measured at $Q_0$, are found to drive the quantum phase transition. The appropriateness of a HMM-like picture to describe the properties of $Ce_{1-x}La_xRu_2Si_2$ indicates that the Fermi-liquid properties, reported by thermodynamic and transport measurements for $x \leq x_c$ (refs 11, 26), may be controlled by antiferromagnetic fluctuations. Indeed, in a HMM-like scenario, the characteristic temperature of the paramagnetic Fermi-liquid regime $T^*$ (Fig. 1) is also the energy scale of the order-parameter fluctuations. As $T_1$ is the energy scale of the antiferromagnetic order-parameter fluctuations in $Ce_{1-x}La_xRu_2Si_2$, we propose that $T_1$ corresponds directly to $T^*$. This correspondence is further supported by the fact that, like $T^*$ in a HMM-like model, $T_1$ decreases in the paramagnetic phase once the quantum phase transition is approached (Fig. 4). The fact that $T_1 = T_N$ for $x = 13$ and 20 % indicates that the parameter $T_1$, which characterizes the low-temperature antiferromagnetic fluctuations, is the relevant magnetic energy scale of both the antiferromagnetic phase and the paramagnetic Fermi-liquid regime. These quantities may be simply related to the RKKY antiferromagnetic exchange, which is the driving force for both antiferromagnetic fluctuations and ordering.

The absence of any anomaly in the local magnetic fluctuations at the quantum phase transition permits us to definitely rule out the CS `local scenario' [13-16] for $Ce_{1-x}La_xRu_2Si_2$. In this scenario, the critical magnetic fluctuations are local (that is, single-site), which implies that the magnetic fluctuations at all wave vectors of the reciprocal lattice have to be maximal at the quantum phase transition. We have shown that this is clearly not the case for $Ce_{1-x}La_xRu_2Si_2$. To validate the CS `local scenario', a prerequisite should be to verify that quantum criticality is a property of the magnetic fluctuations at each momentum transfer Q, that is, by checking that the Q-dependent static susceptibility is, at each Q, diverging or at least maximal at the quantum phase transition. It would be interesting to verify, following a procedure similar to the one introduced here, that the magnetic fluctuations at all wave vectors are maximal at the quantum critical point of the heavy fermion $CeCu_{6-x}Au_x$, which has been presented as the prototype of local magnetic quantum criticality [13-17].

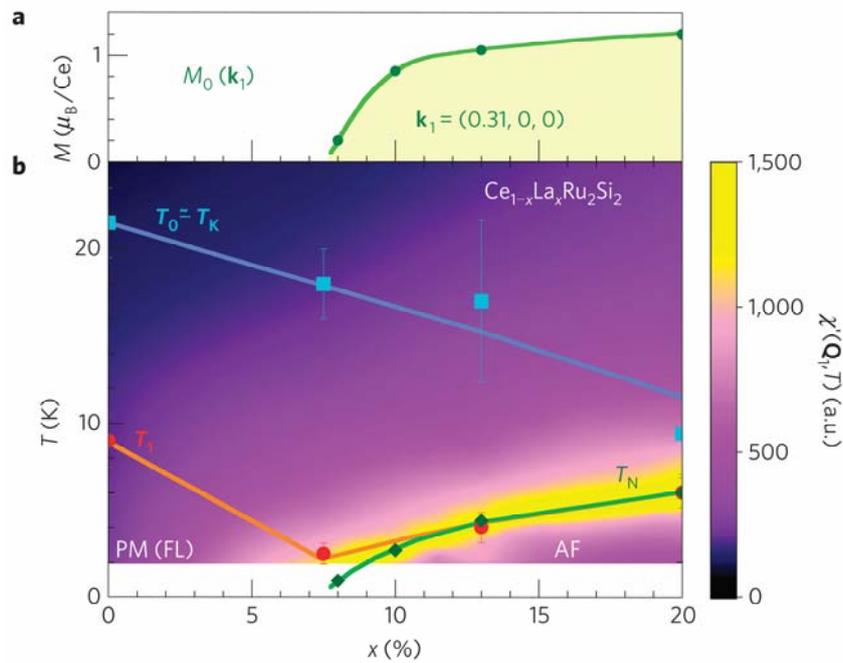

**Figure 4 : *x*-variation of the magnetic order parameter and magnetic phase diagram of $Ce_{1-x}La_xRu_2Si_2$. a**, Variation with La content *x* of the low-temperature order parameter $M_0(\mathbf{k}_1)$ of the antiferromagnetic phase (from ref. 22). **b**, *x*–*T* magnetic phase diagram of $Ce_{1-x}La_xRu_2Si_2$ extracted from neutron scattering experiments. $T_1$ and $T_0$ are determined from the excitation spectra presented here, whereas $T_N$ is taken from earlier diffraction measurements [22] (AF: antiferromagnetic order, PM: paramagnetic regime, FL: Fermi liquid). The lines are guides to the eyes. The intensity of the colour plot corresponds to an extrapolation, in the window $0 \leq x \leq 20$ %, of the staggered static susceptibility $\chi'(\mathbf{Q}_1,T)$ measured here for for $x = 0, 7.5, 13,$ and $20$ %.

The experimental study presented here has revealed that HMM-like models [7-9], based on antiferromagnetic order-parameter fluctuations, are pertinent to describe quantum criticality in the heavy-fermion system $Ce_{1-x}La_xRu_2Si_2$. It has been shown that, although local magnetic fluctuations persist in the antiferromagnetic phase, they are not the driving phenomenon for the quantum phase transition. We mention that, in the studies of $CeRu_2Si_2$-based compounds, the proximity of a valence transition can be rejected because, from the pressure dependence of its magnetic Grüneisen parameter, pure $CeRu_2Si_2$ is expected to become intermediatevalent in the range 2-5 GPa (refs 5, 40). In contrast, the magnetic instability in some heavy-fermion systems may be coupled to a valence instability [5]. In the future, systematic studies of the magnetic fluctuations may be carried out on other heavy-fermion compounds in both the paramagnetic and antiferromagnetic regimes. In the light of the results obtained here for $Ce_{1-x}La_xRu_2Si_2$, further developments of HMM-like magnetic fluctuation theories may be carried out. The possibility of a first-order quantum phase transition, which may be related to a subtle change of the Fermi surface at $x_c$, was omitted in the initial HMM approach and may also be considered (see ref. 41). Furthermore, the consideration of the effects of temperature on the local magnetic fluctuations, in addition to those on the critical antiferromagnetic fluctuations, may be of importance for a proper description of the non-Fermi-liquid regime. A correct microscopic understanding of quantum magnetic criticality may finally permit a better understanding of why superconductivity develops, in many heavy-fermion or strongly correlated electrons systems, in the vicinity of a quantum magnetic instability [42-44].

## Methods

The single crystals studied here were grown by the Czochralsky method. Inelastic neutron scattering experiments were carried out using the triple-axis spectrometers IN12 and IN22 (CRG-CEA) at the Institut Laue Langevin in Grenoble, France, and 4F1 and 4F2 at the Laboratoire Léon Brillouin in Saclay, France. As the neutron intensity is not calibrated, the different sets of data (see Fig. 2a,b) were normalized together at high energy transfers E, where the physics is expected to be independent of x.

## Acknowledgements

We acknowledge B. Vettard, J. Prévitali and J. M. Mignot for support during experiments. We also thank P. Haen, F. Lapierre, B. Fåk, H. Yamagami, M. Lavagna, C. Pépin and M. Continentino for useful discussions.